\documentclass{XrU2005}
\usepackage{epsfig}

\def\arcsec{\hbox{$^{\prime\prime}$}}
\def\arcdeg{\hbox{$^\circ$}}

\def\lae{\mathrel{\raise .4ex\hbox{\rlap{$<$}\lower 1.2ex\hbox{$\sim$}}}}
\def\gae{\mathrel{\raise .4ex\hbox{\rlap{$>$}\lower 1.2ex\hbox{$\sim$}}}}

\def\aap{{A\&A}}

\def\apj{{ApJ}}
\def\apjl{{ApJ}}
\def\apjs{{ApJS}}
\def\mnras{{MNRAS}}

\title{X-ray Emission from the 3C 273 Jet}
\author{H.L. Marshall}
\affil{MIT Kavli Institute, MIT, Cambridge, MA, 02139, USA}
\author{S. Jester}
\affil{Southampton University, Southampton SO17 1BJ, UK}
\author{D.E. Harris}
\affil{Smithsonian Astrophysical Observatory, Cambridge, MA, 02138, USA}
\author{K. Meisenheimer}
\affil{Max-Planck-Institut fur Astronomie, D-69117 Heidelberg, Germany}

\begin{document}


\maketitle

\begin{abstract}

We present results from four recent {\it Chandra} monitoring
observations of the jet in 3C~273 using the ACIS detector, obtained between
November 2003 and July 2004.  We find that the X-ray emission
comes in two components: unresolved knots that are smaller than the
corresponding optically emitting knots and a broad channel that is about
the same width as the optical interknot region.  We compute the jet
speed under the assumption that the X-ray emission is due to inverse
Compton scattering of the cosmic microwave background, finding that the
dimming of the jet X-ray emission to
the jet termination relative to the radio emission may be due to
bulk deceleration.

\end{abstract}

\section{Introduction}

\cite{marshall01} showed that X-rays from the 3C~273 jet follow
the optical emission fairly well.  However, the
X-ray emission is significantly brighter
at the beginning of the jet than at the end while the optical knots
are of similar brightness along the jet.  Optically, the width of the jet is
well resolved at the 0.1\arcsec\ level using the Hubble Space
Telescope (HST) and we find that it is now resolved
in the X-ray band in the cross-jet direction.  Assuming that the X-ray emission
arises from inverse Compton scattering of cosmic
microwave background photons (IC-CMB) \citep{tavecchio,celotti},
we can compute the jet speed for a given (small) angle to
the line of sight that is approximately constant along the jet.

\begin{figure*}
\centering
\includegraphics[width=0.4\linewidth,angle=90]{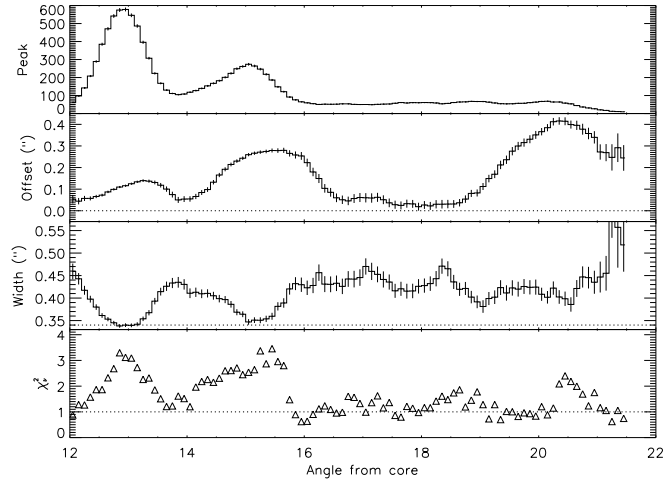}
\caption{Results from fitting Gaussians to the X-ray cross-jet profiles.
From the top, the panels are: Gaussian normalization
(counts per 0.1\arcsec\ bin), angular deviation
from PA=-137.5\arcdeg, the Gaussian dispersion parameter
($\sigma \equiv$ FWHM/2), and 
the reduced $\chi^2$.  The dotted line in the width
panel shows the Gaussian dispersion obtained for an unresolved source. 
Except at 13\arcsec\ and 15.5\arcsec\ from the
core (knots A and B), the jet is resolved in the cross-jet direction.
\label{fig:fitparams}}
\end{figure*}

\section{Cross-Jet Profile}

Results from fits of Gaussians to the X-ray cross-jet profiles are shown
in Fig.~\ref{fig:fitparams}.  The profiles are adaptively binned to
maintain $>15$ counts per bin so there are 7-30 bins per profile.
For comparison, the ACIS readout streak
from the quasar core was fit to a Gaussian, providing a good value for
the Gaussian width,
$\sigma = 0.34$\arcsec\ (or FWHM = 0.80\arcsec), of a point source.
Except in the centers of knots A and B (about 13\arcsec\ and 15.5\arcsec\ from the
core), the jet is resolved.  The reduced $\chi^2$
is near unity
except at the positions of the X-ray bright knots
because the 1D profile
of a point source does not match a Gaussian function.  The residuals of the
Gaussian fits are very similar to those of unresolved sources, confirming
that knots A and B are point-like.
The size of knot A's X-ray emission region
($<$ 0.2\arcsec\ FWHM) is
distinctly smaller than found
optically, where knot A was easily resolved using HST
to be about 0.4\arcsec\ across and 0.7\arcsec\ long \citep{bahcall}.
The physical size differences between optical and X-ray emission regions
are not expected in the IC-CMB model.

For an average observed $\sigma$ of 0.45\arcsec\ and the value for
a point source of 0.35\arcsec, the
inferred intrinsic FWHM of the X-ray emission outside knots A and B is
0.62\arcsec.  This width is comparable to that of the optical emission
between knots \citep{bahcall}.
Thus, the X-ray emission
comes in two components: unresolved knots that are smaller than the
corresponding optically emitting knots and a broad channel that is about
the same size as the optical interknot region.
\section{Jet Speed}

We use the profile of the X-ray emission from the jet, which we compare
to the radio profile to obtain the flow speed along the jet if we accept
the hypothesis that the X-rays are produced by inverse Compton scattering of
microwave background photons in a relativistic jet.
Following \cite{hk02}, \citet{marshall05}
showed that the beaming parameters Ñ- the
cosine of the angle to the line of sight $\cos \theta = \mu$ and
the jet speed $\beta c$ Ñ- are
related to a function, $K$, of the observables.
See \citet{marshall05} for details.  We can solve
their Equation~4 for $\beta$, giving

\begin{equation}
\label{eq:beta}
\beta = \frac{-1-\mu +2 K \mu \pm (1+2 \mu  - 4 K \mu + \mu^2 
	+ 4 K \mu^3)^{1/2}}{2 K \mu^2}
\end{equation}

\noindent
Using the ratio of the X-ray and radio fluxes as a function of
position along the jet and Equation~\ref{eq:beta} above, we compute the
jet $\Gamma = (1-\beta^2)^{(-1/2)}$ upon setting the
angle to the line of sight and using the magnetic field from
\citet{jester02}, which seems to be nearly constant along the
jet.  We find a good
solution for a very small angle to the line of sight, 2.5\arcdeg, where
the jet bulk $\Gamma$ drops from a maximum of 18 down to 2 at the terminal
hotspot.  Thus, the dimming of the jet to
the end relative to the radio emission may be due to bulk deceleration.

\begin{figure}
\centering
\includegraphics[width=0.6\linewidth,angle=90]{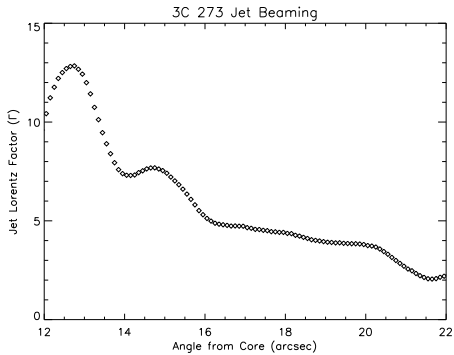}
\caption{Jet beaming factor assuming the IC-CMB model of
the X-ray emission.  The X-ray to radio flux ratio is used as in
the anaysis by \cite{marshall05} but using equation~\ref{eq:beta}
to compute the jet speed and Lorentz factor.  In this model,
the jet dimming results from bulk deceleration.\label{fig:beaming}}
\end{figure}

\section*{Acknowledgments}

This work was supported in part by U.S. Dept. of Energy under contract
No. DE-AC02-76CH03000, Smithsonian Astrophysical Observatory (SAO)
contract SVI-73016 for the Chandra X-Ray Center (CXC), and NASA contract
NAS8-39073.


\begin{thebibliography}{}
\bibitem[Bahcall et al.(1995)]{bahcall}
	Bahcall et al. 1995, \apj, 452, L91
\bibitem[Celotti et al.(2002)]{celotti}
	Celotti, A., Ghisellini, G., and Chiaberge, M. 2001, \mnras, 321, L1
\bibitem[Jester et al.(2002)]{jester02} Jester, S. et al., 2002, \aap, 385, L27
\bibitem[Harris \& Krawczynski(2002)]{hk02} Harris, D.~E.~\& 
	 Krawczynski, H.\ 2002, \apj, 565, 244
\bibitem[Marshall et al.(2001)]{marshall01}
	Marshall, H.L., et al.\ 2001, \apjl, 549, L167
\bibitem[Marshall et al.(2005)]{marshall05}
	Marshall, H.L., et al.\ 2005, \apjs, 156, 13.
\bibitem[Tavecchio et al.(2000)]{tavecchio} Tavecchio, et al.\  2000, \apj, 544, L23

\end{thebibliography}
\end{document}